\documentclass[twocolumn]{article}
\usepackage{graphicx} 
\usepackage{amsmath}
\usepackage{amssymb}
\usepackage{tikz}
\usetikzlibrary{quantikz2}
\usepackage{caption}
\usepackage{subcaption}
\usepackage{authblk}
\usepackage{adjustbox}
\usepackage{algorithm}
\usepackage{algpseudocode}
\usepackage[style=ieee, citestyle=numeric-comp, backend=biber]{biblatex}
\title{Quantum Circuit Training with Growth-Based Architectures}

\author[1,2]{Callum Duffy}
\author[1]{Smit Chaudhary}
\author[1]{Gergana V. Velikova}

\affil[1]{PASQAL, 7 Rue Léonard de Vinci, Massy, France, 93100}
\affil[2]{Centre for Data Intensive Science and Industry (DISI), University College London, Gower Street, London WC1E 6BT, United Kingdom}

\date{}

\begin{document}

\maketitle

\begin{abstract}

This study introduces growth-based training strategies that incrementally increase parameterized quantum circuit (PQC) depth during training, mitigating overfitting and managing model complexity dynamically. We develop three distinct methods: Block Growth, Sequential Feature Map Growth, and Interleave Feature Map Growth, which add reuploader blocks to PQCs adaptively, expanding the accessible frequency spectrum of the model in response to training needs. This approach enables PQCs to achieve more stable convergence and generalization, even in noisy settings. We evaluate our methods on regression tasks and the 2D Laplace equation, demonstrating that dynamic growth methods outperform traditional, fixed-depth approaches, achieving lower final losses and reduced variance between runs. These findings underscore the potential of growth-based PQCs for quantum scientific machine learning (QSciML) applications, where balancing expressivity and stability is essential. 
\end{abstract}

\section{Introduction}

Parameterized Quantum Circuits (PQCs) are promising machine learning models within quantum computing, offering potential applications across optimization \cite{farhi2014quantumapproximateoptimizationalgorithm, verdon2019learninglearnquantumneural,Hadfield_2019,PhysRevA.97.022304,Streif_2020}, scientific computing \cite{Kyriienko_2021}, and other complex tasks \cite{Preskill2018quantumcomputingin}. PQCs provide unique expressivity that may surpass classical models by harnessing quantum features like entanglement and superposition. However, deploying PQCs on noisy intermediate-scale quantum (NISQ) devices introduces significant challenges. Device limitations lead to noise accumulation, and as circuit depth increases, PQCs are susceptible to barren plateaus—regions in the optimization landscape where gradients vanish exponentially with the number of qubits, severely hindering effective training \cite{McClean2018BarrenPI}.\newline 

The initialization and architecture of PQCs significantly impact model convergence and performance. Some initialization techniques include initializing sets of gates to identity matrices \cite{Grant_2019}, drawing trainable parameters from specific Gaussian distributions \cite{zhang2022escapingbarrenplateaugaussian}, or using small rotation angles for single-qubit gates \cite{Holmes_2022}. These techniques have demonstrated marked improvements over naïve random initialization and emphasize the importance of parameter selection in PQCs.\newline

Additionally, the chosen architecture of a quantum neural network (QNN) shapes its inductive biases, convergence speed, and generalization capabilities. The noisy nature of current quantum devices limits the feasible depth of PQCs, highlighting the need for compact architectures with sufficient expressibility to solve tasks. Determining the optimal architecture a priori is often intractable, which has motivated the development of adaptive methods to adjust model complexity dynamically during training. Classical neural networks have successfully used dynamic models to achieve compactness \cite{Evci2022GradMaxGN}. In contrast, quantum approaches such as neural architecture search \cite{duong2022quantum, Du_2022}, evolutionary \cite{lourensHierarchicalQuantumCircuit2023}, and adaptive methods \cite{PRXQuantum.2.020310} have shown promise in finding efficient architectures without requiring fully parameterized deep circuits from the start of training.\newline 

In this work, we combine parameter initialization and adaptive growth strategies to develop compact PQCs that enhance convergence, improve training stability, and generalize effectively, even in noisy environments. We propose growth-based training strategies to incrementally increase circuit depth during training, specifically focusing on QNNs structured as reuploader circuits. This growth approach considers reuploader circuits as truncated Fourier series \cite{Schuld_2021}, whereby we add feature map gates, gradually expanding the frequency spectrum available to the model. By selectively adding parameterized gates, our methods dynamically manage model capacity, reducing overfitting and adjusting the model’s expressivity in response to the complexity of the target function. We introduce three distinct growth methods which incrementally add reuploader blocks during training. Prior work in Ref. \cite{Skolik_2021} shares some parallels to our work where PQCs grow layer-by-layer with new blocks initialized to zero, our method differs both in the exact way we grow PQCs, and initialization strategy since we use identity-initialized blocks.\newline

We evaluate the proposed methods through a series of tasks, including learning the output of randomly initialized PQCs and solving the 2D Laplace equation using a notable QSciML method, which uses quantum devices with differentiable quantum circuits (DQCs) \cite{Kyriienko_2021}. Additionally, we examine the regularizing properties of growing PQCs in noisy scenarios, demonstrating their ability to adapt model complexity to data characteristics, thus effectively balancing bias and variance. The results underscore the advantage of growing circuit strategies in achieving lower final losses and greater robustness compared to traditional methods, which train all PQC parameters from the outset.\newline

\section{Methods}\label{sec:2}

Reuploader models contain two sets of unitaries, those that encode input data $\Vec{x}$ into feature map unitaries $\hat{F}(\vec{x})$ and ansatz unitaries $\hat{U}(\Vec{\theta})$ which contain trainable parameters $\Vec{\theta}$. To then form a reuploader model products of $\hat{F}(\vec{x})\hat{U}(\vec{\theta})$ are repeated a user-specified $L$ times. More broadly, we can define a feature map as a tensor product
\begin{equation}
    \hat{F}(\Vec{x}) = \bigotimes_m e^{-\frac{i}{2}\hat{G}_m(\gamma_m)\phi(\Vec{x})},  
\end{equation}

of $m$ feature map gates with generator Hamiltonian's $\hat{G}_m(\gamma_m)$, which depend on $\gamma_m$ which are non-trainable and $\phi(\Vec{x})$ some function encoding $\Vec{x}$. When the PQC consists of alternating unitary blocks of feature maps $\hat{F}(\Vec{x})$ and ansatzs $\hat{U}(\Vec{\theta})$, thus taking the form of a reuploader circuit, upon measurement the output takes the form of a truncated Fourier series given by
\begin{equation}
    f(\Vec{x}, \Vec{\theta}) = \sum_{\omega_j \in \Omega} \Vec{c_j}(\Vec{\theta})e^{i\vec{\omega}_j\cdot\phi(\Vec{x})},
\end{equation}

where $\Vec{c_j}$ are the coefficients of the Fourier mode with frequencies $\vec{\omega}_j$ \cite{Schuld_2021}. The accessible frequencies of the model $\Omega$ are determined by the eigenspectrum of $\hat{G}_m$. Given the $\hat{G}_m$ generators commute we can write $\hat{G}=\sum_m\hat{G}_m$, then $\Omega$ contains the gaps in the eigenspectrum of $\hat{G}$. The size of the spectrum $K=|\Omega|$ depends on the number of repetitions of the encoding gates $\hat{F}(\vec{x})$, where there is a linear dependence on the size $K$ of the spectrum with respect to the number of repetitions $L$ of the encoding gates. To equate this to the commonly used angle-encoding \cite{PhysRevA.102.032420}, $\hat{G}$ is a single qubit operator such that $\hat{G} = \sum_{m=1}^{N} = \gamma_m \hat{P}^m /2$ where $N$ is the number of qubits and $\hat{P}^m$ is a chosen Pauli matrix applied to qubit $m$ as a series of tensor products, $\gamma_m = 1$, $\phi(\Vec{x})=\vec{x}$. By this construction, the frequency spectrum of the model is fixed, as such only the coefficients $\Vec{c_j}$ of the function can be altered during training. The work in Ref \cite{PhysRevA.109.042421}, recognized this and introduced a new feature map generator that contained trainable parameters $\vec{\psi}$, allowing for the frequency spectrum of the model to alter during training since the eigenspectrum of the generator can now change $\hat{G}_m(\gamma_m, \vec{\psi})$. The output from a PQC in this case can be stated as
\begin{equation}
    f(\Vec{x}, \Vec{\theta}, \Vec{\psi}) = \sum_{\vec{\omega}_j\in\Omega} \vec{c}_j(\vec{\theta},\hat{C})e^{i\vec{\omega}_j(\vec{\psi})\cdot\phi(\vec{x})},
\label{eq:tffm}
\end{equation}

where $\hat{C}$ refers to a measurement operator. For this study, we exclusively study models which output functions of the form in Equation \ref{eq:tffm}, for their spectral richness and greater flexibility in how models can be initialized. Specifically, $\hat{G}$ takes the form, $\hat{G}=\sum_{m=1}^{N}\psi_m\hat{X}^m/2$, where we have chosen the single qubit Pauli-X operator $\hat{X}$ and we have set $\gamma_m=1$.\newline

Using this form of reuploader circuit, we now present the methods we devised for dynamically increasing the depth of a PQC during training. However, before we do so these methods share some common fundamental components. First, one must decide when to grow the circuit. We choose to let circuits grow after a predefined number of epochs. However, a more adaptive approach could involve monitoring metrics such as performance on a validation set, with growth triggered when improvement during training plateaus. Second, the decision of by how much to grow a circuit is determined by the number of unitary gates added during each growth stage. This quantity is a hyperparameter that may depend on the specific problem requirements. When adding gates during training, it is crucial to preserve the function currently represented by the model. To achieve this, the added gates are initialized as identity gates. We accomplish this by introducing gates in pairs: the first gate in each pair is parameterized by drawing values from a selected probability distribution, while the parameters of the second gate are set to cancel the rotation produced by the first. Initializing new gates as identity allows for non-zero gradients on these parameters from the start of gradient descent \cite{Grant2019initialization}, enabling more effective parameter updates during early stages of training.\newline 

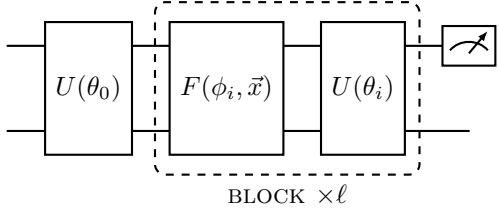
\begin{figure}
\centering
\begin{quantikz}
& \gate[2]{U(\theta_0)} &
\gate[2]{F(\phi_i, \vec{x})}\gategroup[2,steps=2,style={dashed,rounded
corners, inner
xsep=2pt},background,label style={label
position=below,anchor=north,yshift=-0.2cm}]{{\sc
block $\times \ell$}} & \gate[2]{U(\theta_i)} & \meter{} \\
& & & &  
\end{quantikz}
\caption{PQC growing method block growth where both feature maps and ansatzs blocks are added during training, with some specified block being added $\ell$ times every time the PQC grows.}
\label{circ:block_grow}
\end{figure}

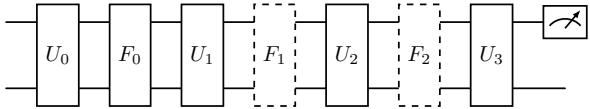
\begin{figure}
\centering 
\begin{adjustbox}{width=0.48\textwidth}
\begin{quantikz}
& \gate[2]{U_0} &
\gate[2]{F_0} & \gate[2]{U_1} &
\gate[2, style={dashed}]{F_1} & \gate[2]{U_2} & 
\gate[2, style={dashed}]{F_2} & \gate[2]{U_3} & \meter{} \\
& & & & & & & & 
\end{quantikz}
\end{adjustbox}
\caption{Sequential feature map growth, where feature map gates are added to the PQC when it grows, adding these gates from left to right in between existing ansatz gates.}
\label{circ:sequential_fm_grow}
\end{figure}

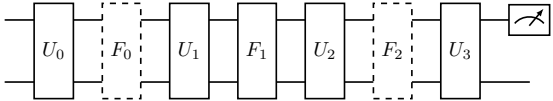
\begin{figure}
\centering
\begin{adjustbox}{width=0.45\textwidth}
\begin{quantikz}
& \gate[2]{U_0} &
\gate[2, style={dashed}]{F_0} & \gate[2]{U_1} &
\gate[2]{F_1} & \gate[2]{U_2} & 
\gate[2, style={dashed}]{F_2} & \gate[2]{U_3} & \meter{} \\
& & & & & & & & 
\end{quantikz}
\end{adjustbox}
\caption{Interleave feature map growth, where feature map gates are added to the PQC when it grows, adding these gates from the middle out in between existing ansatz gates.}
\label{circ:interleave_fm_grow}
\end{figure}

We propose three specific methods for growing PQCs:
\begin{itemize}
    \item \textbf{Block Growth}: In this method, block growth depicted in Figure \ref{circ:block_grow}, the circuit depth is progressively increased by appending predefined blocks of unitaries when greater expressivity is needed. Each added block follows a typical reuploader structure, containing both feature map and ansatz gates. At each growth stage, a user-defined number, $\ell$, of these blocks are appended to the circuit. This addition increases the size of the frequency spectrum $\Omega$ (due to added feature maps) and expands the range of accessible Fourier coefficients $c_n$ (due to added ansatz blocks), as outlined in Equation \ref{eq:tffm}.

    \item \textbf{Seq FM Growth}: In this approach Sequential feature map growth (Seq FM Growth) depicted in Figure \ref{circ:sequential_fm_grow}, all ansatz blocks are included from the start of training, but only a limited number of feature map blocks are initially present. As training proceeds, feature map blocks are sequentially added between ansatz blocks, proceeding from left to right across the circuit. This approach increases the model’s frequency range incrementally without altering the expressivity of the Fourier coefficients.

    \item \textbf{Int FM Growth}: This method, interleave feature map growth (Int FM Growth) shown in Figure \ref{circ:interleave_fm_grow}, follows a similar structure to sequential feature map growth, but feature map blocks are inserted from the center outward in-between ansatz blocks. This interleaved structure modifies the accessible frequency range without impacting the expressivity of the Fourier coefficients, providing an alternative strategy for frequency expansion within the model.
\end{itemize}

Each method offers a distinct pathway for enhancing the model's expressive power through frequency range expansion or increased Fourier coefficient variety, supporting adaptive, structured growth in PQCs during training. We summarize the training procedure for a growing PQC in pseudocode in Algorithm \ref{alg:growth_algo} \footnote{We also provide the code used to implement these techniques https://github.com/callumfduffy/QGrow}. 

\begin{algorithm}
\caption{Pseudocode for growing circuit.}\label{alg:growth_algo}
\begin{algorithmic}
\State Initialize QNN model $M$ with parameters $\vec{\theta}$, $\vec{\psi}$
\State Define optimizer $Opt$ and training hyperparameters: epochs $E$, loss function $L$, criteria to grow $G$  
\State Load data: data $=\{x_{train}, y_{train}, x_{test}, y_{test}\}$
\Procedure{Train}{$M, Opt, E, G, L$, data}
\For{$i \gets 1$ to $E$}
    \State \textbf{Forward pass:} {$\vec{y} \gets M(\vec{x}_{train})$}  
    \State \textbf{Compute loss:} {$l \gets L(\vec{y}, \vec{y}_{train})$}
    \State \textbf{Update parameters:} {$\vec{\theta},\vec{\phi} \gets Opt(\vec{\theta}, \vec{\phi}, l) $} 
    \State \textbf{Evaluate M:} {$l_{test} \gets L(M(\vec{x}_{test}), \vec{y}_{test})$}
    
    \If{$G$ true}  \Comment{if $M$ met criteria to grow} 
        \State \textbf{Grow M} \Comment{e.g. using block growth}
    \EndIf
\EndFor
\EndProcedure
\end{algorithmic}
\end{algorithm}
\section{Results}

\subsection{Student-Teacher}\label{sec:3.1}
\begin{figure*}
\centering
\begin{adjustbox}{width=0.9\textwidth}
    \begin{quantikz}
        & \gate{R_y(\theta_{10})} &  \gate{R_y(\theta_{11})} & \ctrl{1} & \gate{R_x(\phi_{00} x_1)}\gategroup[2,steps=5,style={inner
        sep=6pt}]{repeat $\times L$} & \gate{R_x(\phi_{01} x_1)} & \gate{R_y(\theta_{00})} & \gate{R_y(\theta_{01})} & \ctrl{1} & \meter{}\\
        & \gate{R_y(\theta_{10})} &  \gate{R_y(\theta_{11})} & \targ{} & \gate{R_x(\phi_{10} x_2)} &  \gate{R_x(\phi_{11} x_2)} & \gate{R_y(\theta_{10})} &  \gate{R_y(\theta_{11})} & \targ{} & 
    \end{quantikz}
\end{adjustbox}
    \caption{Teacher circuit architecture for $2$ qubits, randomly initialized to create a dataset for PQCs learn the output of which also follow this circuit architecture.}
    \label{circ:student_teacher_circuit}
\end{figure*}
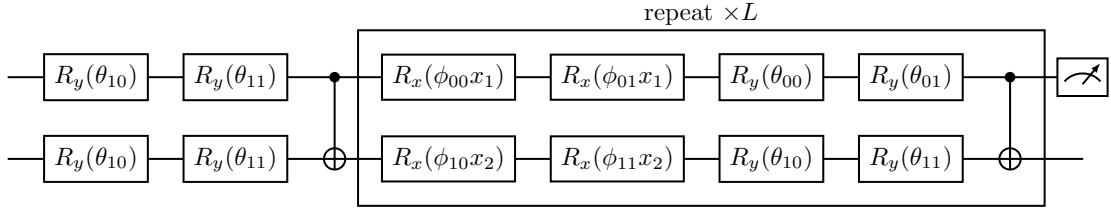
To evaluate our proposed training methods, we tackle the problem of learning the output of a randomly initialized PQC, where the models we train have the same underlying circuit structure. The PQC we wish to learn the output of has a reuploader structure as seen in Figure \ref{circ:student_teacher_circuit}, the models we train have the same matching reuploader blocks but with blocks repeated $\ell$ times which may or may not have the same number of repetitions $L$ as the circuit we wish to learn the output of. This setup allows us to focus exclusively on optimizing parameters, as the optimal solution is guaranteed to lie within the model's solution space when $\ell = L$. This approach serves as a form of student-teacher problem, where the teacher circuit provides the output that the student models attempt to learn, given identical reuploader block structures.\newline 

The target circuit, shown in Figure \ref{circ:student_teacher_circuit}, consists of $R_x$ and $R_y$ rotations, with a single Pauli-Z measurement on the first qubit. Each gate type is repeated twice, as described previously. To learn the output of this fixed, randomly initialized circuit of depth $L$, we explore several training methods. These methods include dynamic PQC growth techniques (detailed in Section \ref{sec:2}) as well as a baselines where the circuit is static, where all gates are present and optimized from the outset, we call this method complete-depth learning (CDL).\newline 

The CDL method is implemented in two configurations: one with gates randomly initialized from a uniform distribution which we denote as RAND, and the other initialized to an identity circuit labeled as $\mathbb{I}$ using the same scheme as the growing circuits. Additionally, we test CDL circuits with varying numbers of reuploader layers $\ell$, which either match the teacher circuit depth $\ell=L$ or exceed it $\ell>L$ for more expressive configurations. We label all the PQCs we trained as follows:\newline 

\begin{itemize}
    \item Block growth
    \item Seq FM growth 
    \item Int FM growth
    \item CDL $\ell=L$ (RAND)
    \item CDL $\ell>L$ (RAND)
    \item CDL $\ell=L$ ($\mathbb{I}$)
    \item CDL $\ell>L$ ($\mathbb{I}$)
\end{itemize}

Each PQC type was evaluated over 50 random seeds. The best loss achieved during each run is reported and shown with boxplots in Figures \ref{fig:boxplot_1d_student_teacher} and \ref{fig:boxplot_2d_student_teacher}, while tables \ref{table:student_teacher_1d} and \ref{table:student_teacher_2d} summarize the mean, standard deviation, and the best and worst achieved losses across all seeds which can be found in appendices \ref{appx:11}, \ref{appx:12} respectively. The loss function of choice for these experiments was the mean squared error (MSE), with all models trained for $1000$ epochs. For each problem we generated input data $\vec{x}$ over the domain $\vec{x}\in[0,2\pi]$, selecting $500$ random points for training and an additional $500$ points for testing.\newline

\subsubsection{1-qubit}

\begin{figure*}
    \centering
    \begin{subfigure}{0.45\textwidth}
        \centering
        \includegraphics[width=\textwidth]{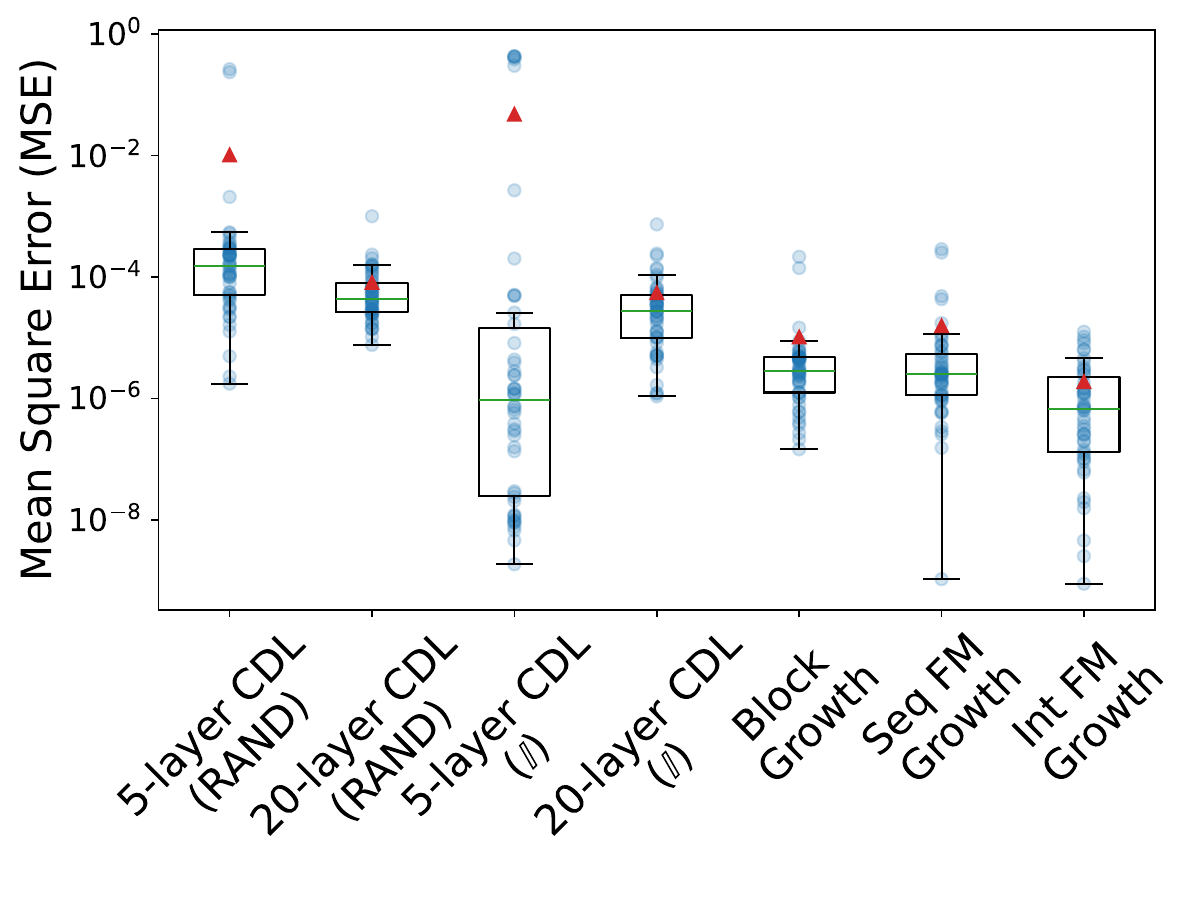}
        \caption{1-qubit}
        \label{fig:boxplot_1d_student_teacher}
    \end{subfigure}
    \hfill
    \begin{subfigure}{0.45\textwidth}
        \centering
        \includegraphics[width=\textwidth]{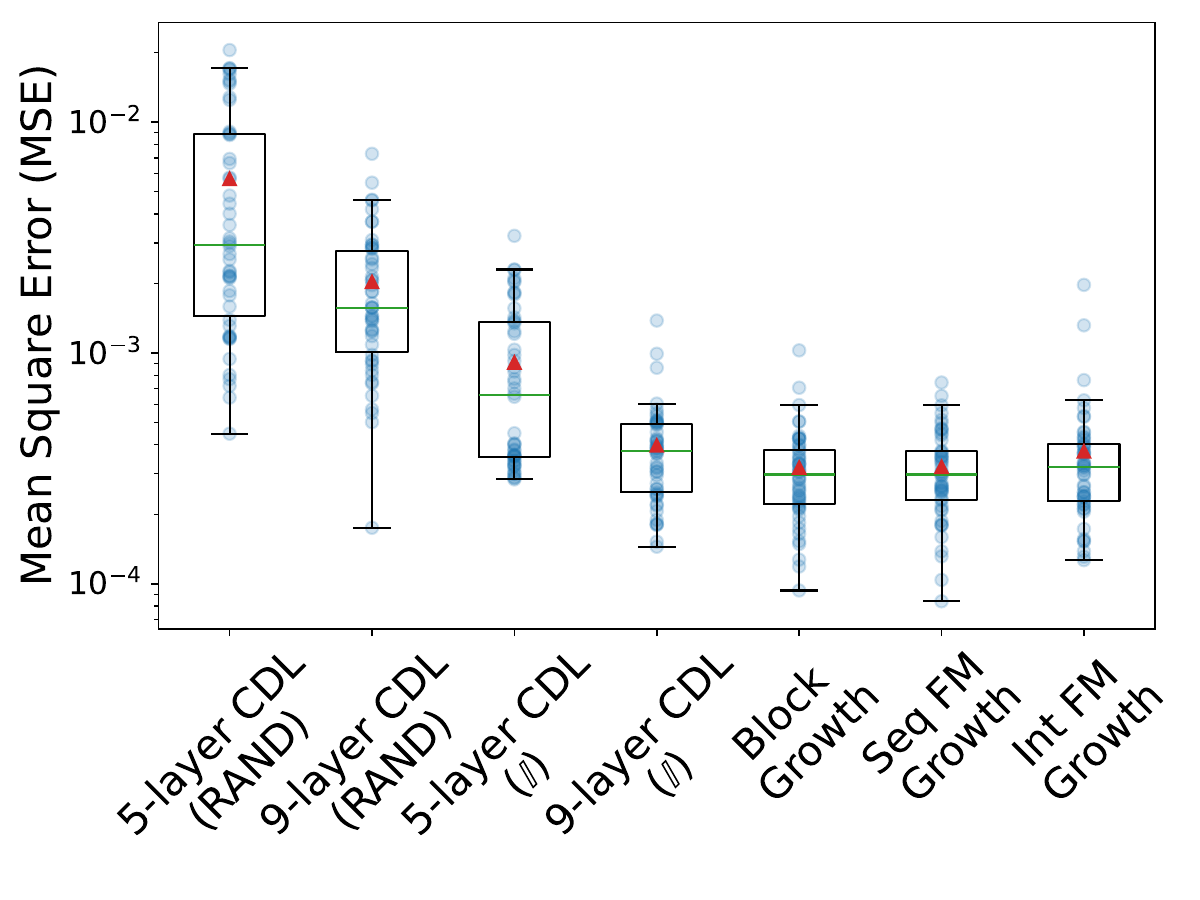}
        \caption{2-qubit}
        \label{fig:boxplot_2d_student_teacher}
    \end{subfigure}
    \caption{Mean squared error performance for different training methods on the student-teacher regression tasks. Each boxplot shows the distribution of MSE values across $50$ random seeds for seven models: 5-layer CDL (RAND), 20-layer CDL (RAND), 5-layer CDL (identity initialization), 20-layer CDL (identity initialization), Block Growth, Sequential Feature Map Growth, and Interleave Feature Map Growth. The blue dots represent the MSE values for each of the runs, green horizontal lines represent the median, red triangles the mean.}
    \label{fig:student_teacher}
\end{figure*}

For the single-qubit circuit, the teacher circuit parameters were randomly initialized, with ansatz gate parameters sampled from a uniform distribution between $0.0$ and $0.1$, and feature map parameters between $0$ and $\pi/9$. All models trained on this dataset used the Adam optimizer with a learning rate of $0.1$.\newline

Figure \ref{fig:boxplot_1d_student_teacher} and table \ref{table:student_teacher_1d} show the performance of the different models on this single-qubit dataset. The poorest performance was observed in the randomly initialized CDL models with $5$ and $20$ layers, despite the $5$-layer model matching the teacher circuit's depth and structure. This suggests that random initialization with a wide parameter range (i.e., between $0$ and $\pi$) may hinder convergence even in smaller models.\newline 

The identity-initialized CDL models showed mixed results. For the $20$-layer model, identity initialization yielded minimal improvement, while for the $5$-layer model, results were more promising. In this case, many runs achieved substantially better performance than their randomly initialized counterparts, although the variance in the best losses remained relatively high across runs.\newline 

Among the dynamically growing circuit methods, block growth, which expanded to $5$ layers, demonstrated more consistent performance by reducing the spread in best losses and improving average-case outcomes. The Int FM growth method outperformed all other models, achieving the lowest mean, best-case, and worst-case MSE among the tested methods, followed closely by the Seq FM growth model, which also demonstrated competitive performance but with slightly higher variance.\newline 


\subsubsection{2-qubit}

In the 2-qubit case, the teacher circuit consisted of $5$ reuploader layers following the structure shown in Figure \ref{circ:student_teacher_circuit}. To initialize the teacher circuit the parameters of the ansatz and feature map gates, we sampled values uniformly between $0.0$ and $\pi/5$. The results of learning this circuit’s output are displayed in Figure \ref{fig:boxplot_2d_student_teacher} and table \ref{table:student_teacher_2d}.\newline 

As with the 1-qubit case, a similar trend emerges. However, we note that the identity-initialized 5-layer CDL model is no longer competitive with the best-case results achieved by the growing circuit methods. Over the course of training, each of the growing circuits reached a depth of 5 layers, and all three growth methods exhibited similar performance across evaluation metrics. Notably, each growing model consistently achieved an order of magnitude improvement in MSE compared to the randomly initialized CDL models across all metrics. Additionally, the growing models outperformed both the identity-initialized CDL circuits.\newline 

Across both the $1$-qubit and $2$-qubit studies, the results suggest that the growing circuit methods provide a robust training approach for PQCs. In both cases, these methods yielded lower MSE in both average and best-case scenarios compared to static, CDL training schemes, highlighting their effectiveness in achieving more optimal parameter configurations. This indicates that incremental growth during training can enhance convergence and model performance, particularly as circuit complexity increases. Furthermore, the consistently lower variance observed with the growing methods suggests that these techniques may provide greater training stability, potentially making them preferable in applications where reliable convergence is essential.\newline

\subsection{Noisy Student-Teacher}

\begin{figure}
\centering
    \includegraphics[width=.45\textwidth]{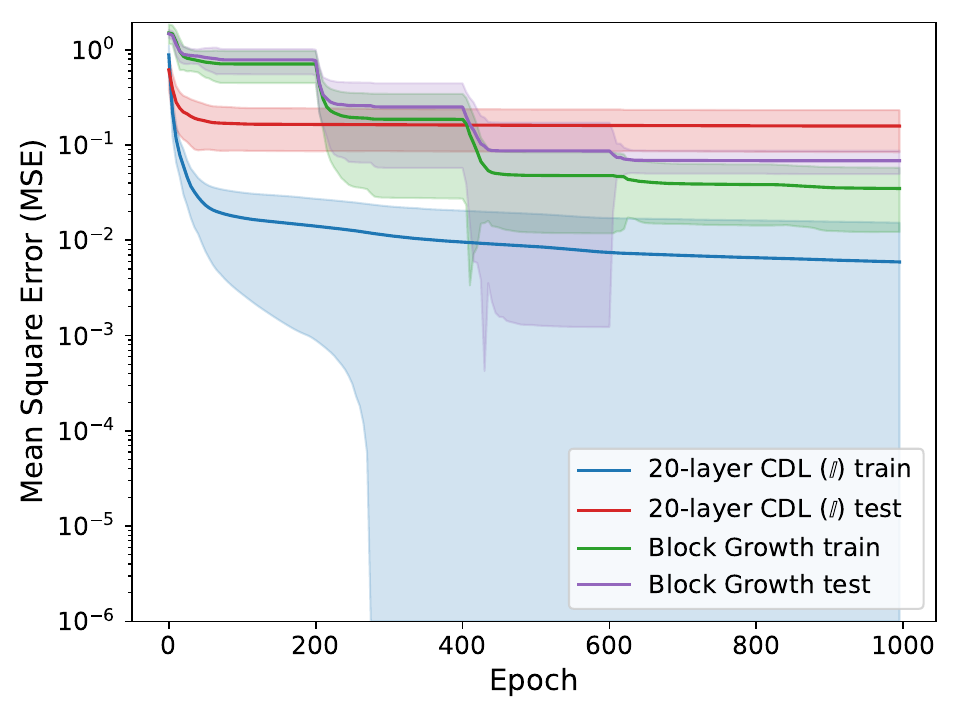}
    \caption{The mean square error of both the $20$-layer CDL and block growth models on the $1$D noisy student-teacher model on a log-scale over the course of training for $1000$ epochs. Reporting the mean train and test error as the solid lines and the standard deviation on the mean as the shaded regions.}
    \label{fig:noisy_student_teacher_losses.pdf}
\end{figure}

\begin{figure*}[t]
  \centering
  \begin{subfigure}[t]{.245\textwidth}
    \centering
    \includegraphics[width=\linewidth]{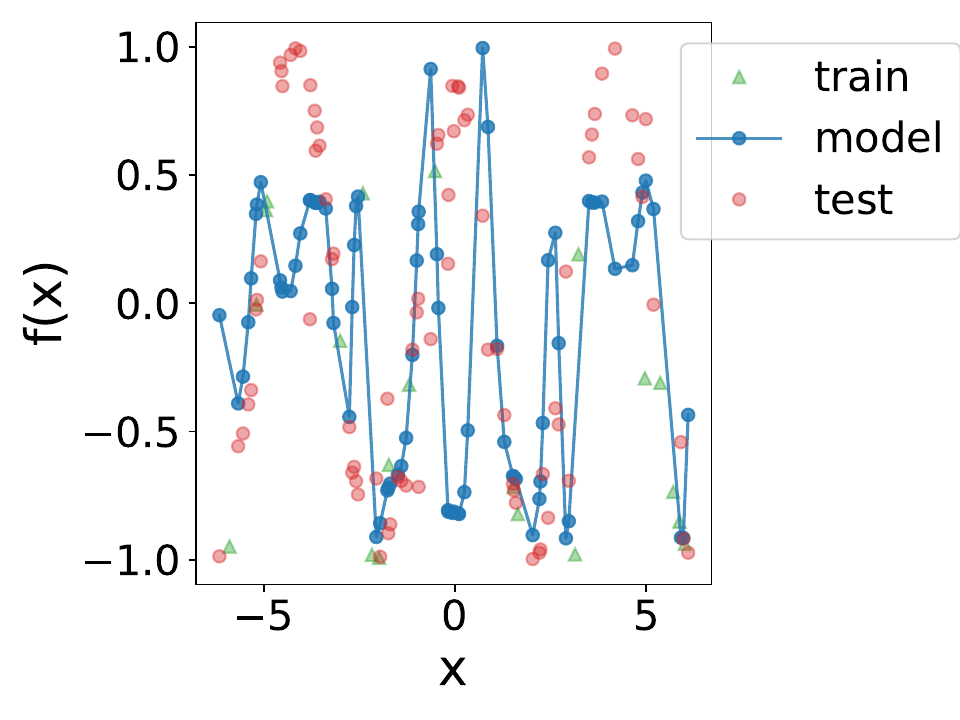}
    \caption{Worst $20$-layer CDL}
  \end{subfigure}
  \hfill
  \begin{subfigure}[t]{.245\textwidth}
    \centering
    \includegraphics[width=\linewidth]{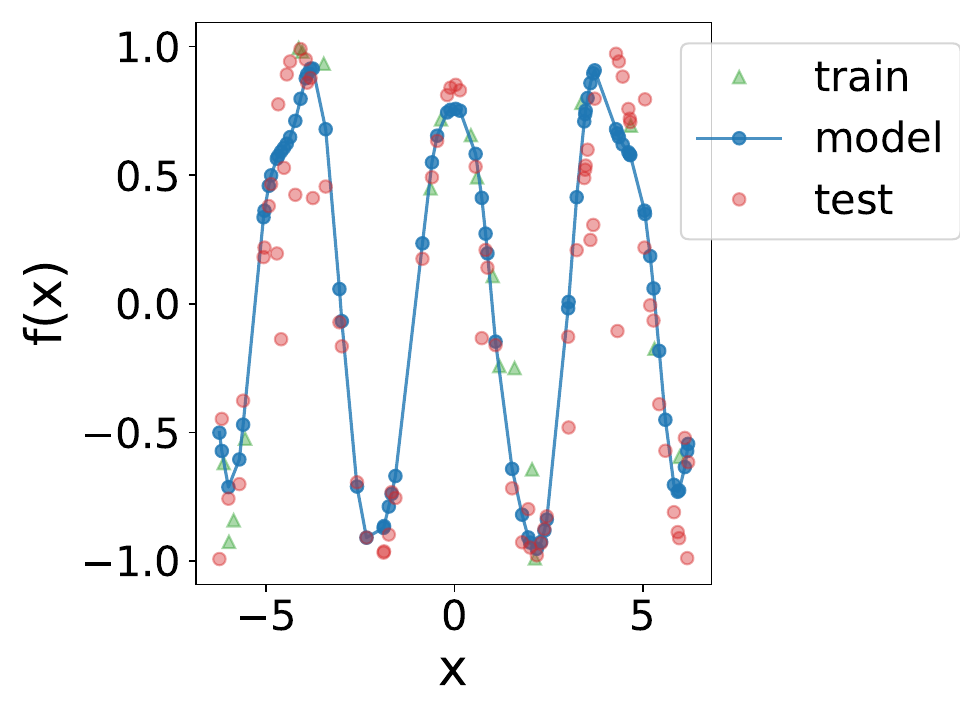}
    \caption{Best $20$-layer CDL}
  \end{subfigure}
  \hfill
  \begin{subfigure}[t]{.245\textwidth}
    \centering
    \includegraphics[width=\linewidth]{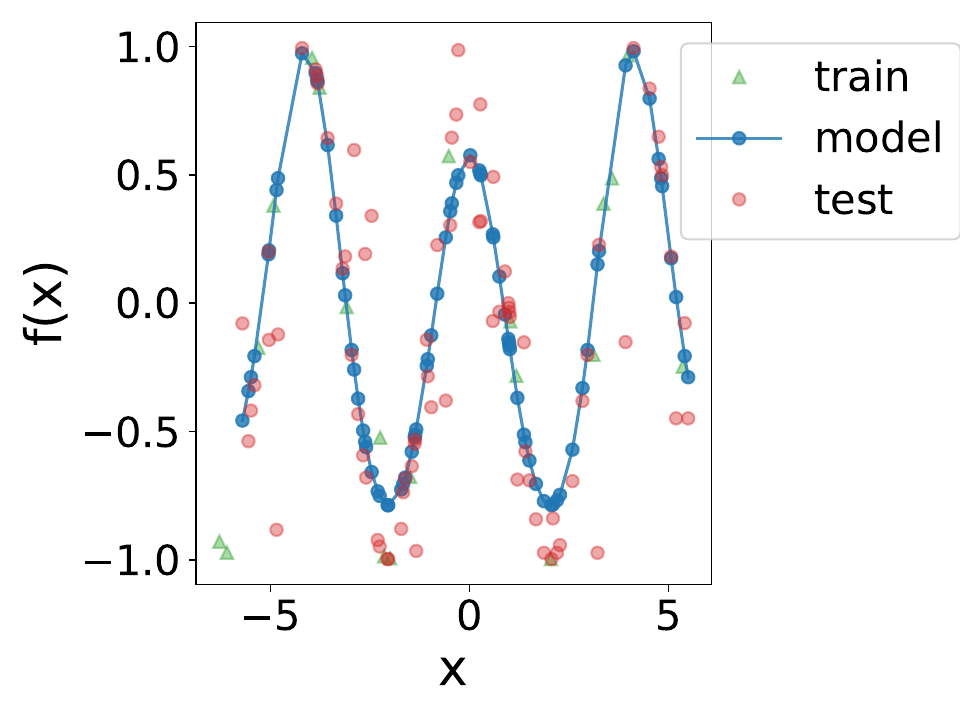}
    \caption{Worst block growth}
  \end{subfigure}
  \hfill
  \begin{subfigure}[t]{.245\textwidth}
    \centering
    \includegraphics[width=\linewidth]{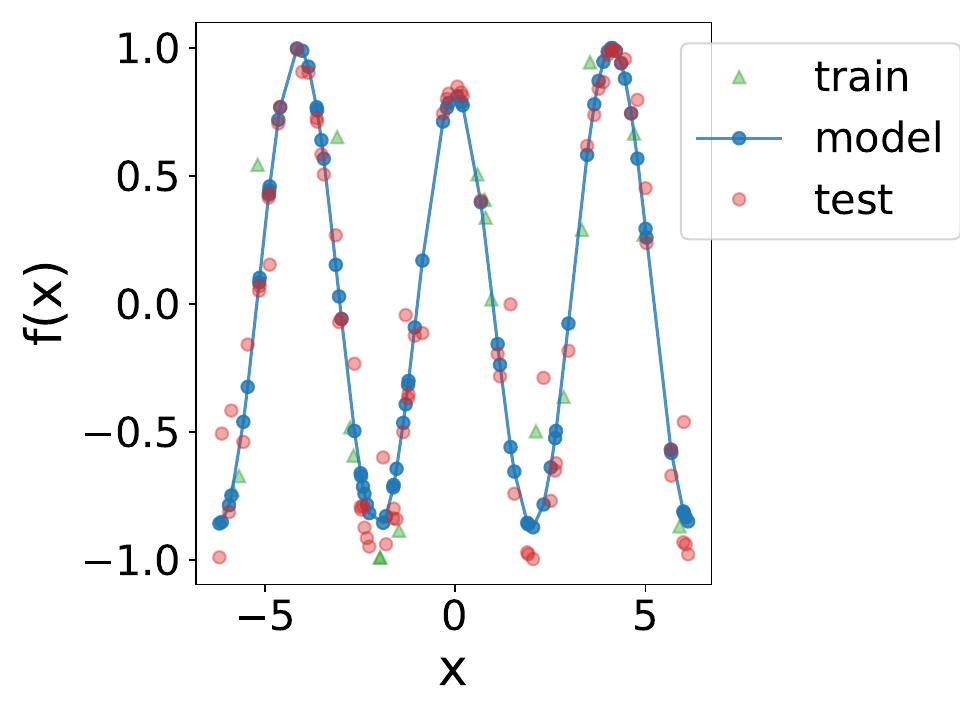}
    \caption{Best block growth}
  \end{subfigure}
\caption{The outputs of PQCs trained on the noisy $1$D student-teacher dataset. Depicting both $20$-layer CDL and block growth models, specifically showing the outputs of the best and worst performing seeds for each.}
  \label{fig:noisy_best_worst}
\end{figure*}

We now conduct a study to examine the regularizing effects of dynamically adding gates to a PQC during training, thereby adaptively controlling the model's complexity. This experiment can be viewed as exploring how a growing quantum circuit impacts the bias-variance trade-off, a fundamental aspect in managing overfitting. To investigate this, we perform a regression task similar to that in Section \ref{sec:3.1}, but with added noise to simulate a more challenging scenario.\newline 

In many natural settings, low-frequency components dominate data, as observed in natural images \cite{fridovichkeil2022spectralbiaspracticerole} and in physical systems governed by partial differential equations (PDEs), such as fluid flows, where energy decays with frequency. This results in lower-frequency components carrying larger magnitudes than high-frequency ones \cite{george2024incrementalspatialspectrallearning}. Models with unrestricted access to higher frequency modes are thus more prone to overfitting, as they may interpret noise or other high-frequency components as meaningful data. In this experiment, we aim to demonstrate that by gradually expanding the accessible frequency spectrum of a PQC, we can effectively regularize the model, mitigating the risk of overfitting.\newline 

The setup is as follows: we use the same circuit structure as before, but with noise added to the teacher circuit’s outputs, sampled from a normal distribution with mean $\mu=0$ and standard deviation $\sigma=0.5$. We reduce the training dataset to $20$ points to emphasize overfitting tendencies in deeper circuits that access multiple frequency modes, as they may interpret the noise as a high-frequency signal. For testing, we sample $80$ points to evaluate generalization.\newline 

The two models we compare are a block-growth model and a $20$-layer CDL PQC, both following the structure in Section \ref{sec:3.1} and initialized to identity blocks. Figure \ref{fig:noisy_student_teacher_losses.pdf} shows the training and test loss curves over $50$ seeds for each model, with mean and standard deviation on the losses reported. The $20$-layer CDL circuit, as expected, exhibits overfitting: the training loss curve achieves a final average of $0.005$, while the test loss curve ends at $0.158$, revealing a significant gap between training and test performance.\newline 

In contrast, the block-growth model, which incrementally increases the circuit depth until no further improvement is observed, mitigates overfitting effectively. Here, the train and test losses align closely throughout training, with final average losses of $0.035$ (train) and $0.063$ (test)—a significant improvement compared to the CDL model. These values are of the same order of magnitude, whereas the $20$-layer CDL model’s losses differ by two orders of magnitude, underscoring the regularizing benefit of the growing circuit approach.\newline 

We further illustrate this effect by presenting the best and worst-performing models on the dataset in Figure \ref{fig:noisy_best_worst}, which demonstrates that the growing circuit method is not only more robust, with less variation across random seeds, but also better captures the underlying pattern in the data, avoiding the misleading influence of noise.\newline 

\subsection{2D Laplace}

\begin{figure}[t]
\centering
    \includegraphics[width=.45\textwidth]{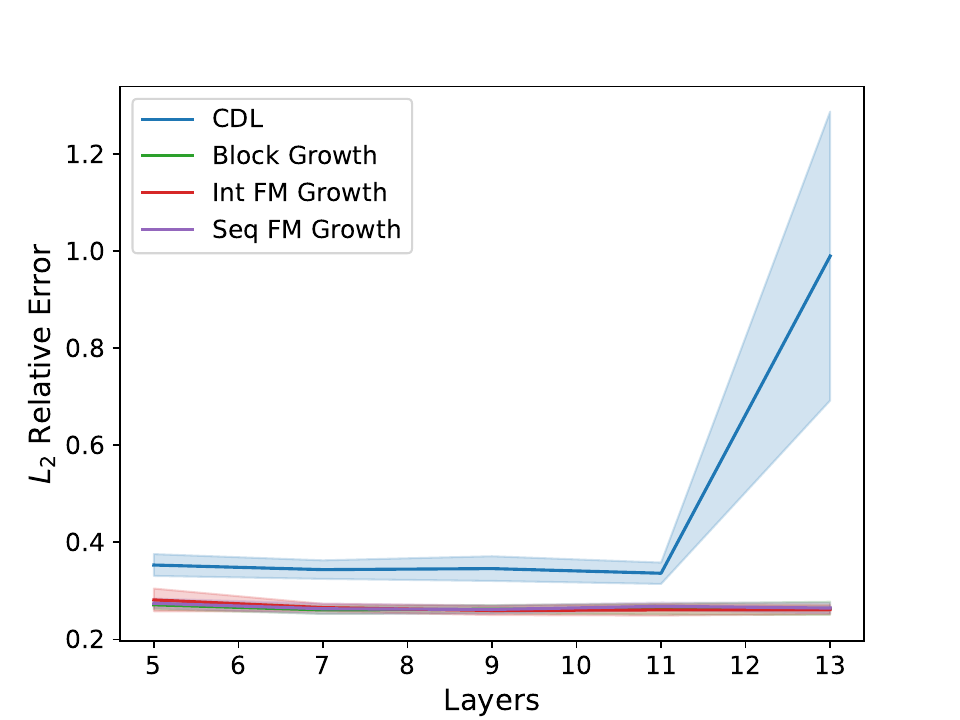}
    \caption{$L_2$ relative error of various models solving the $2$D Laplace equation. The plot compares the mean $L_2$ relative error in bold with standard error shaded for each model and how many reuploader layers it began with or grew to over $50$ training runs. Models include complete-depth learning (CDL) approaches with identity ($\mathbb{I}$) initialization, alongside growth-based methods (Block Growth, Sequential Feature Map Growth, and Interleave Feature Map Growth).}
    \label{fig:boxplot_2d_laplace}
\end{figure}

We now evaluate the effectiveness of the growing PQC method on a more complex, practically relevant problem: solving the $2$D Laplace equation. This equation is a partial differential equation (PDE) describing various steady-state physical systems and is given by:
\begin{equation}
    \dfrac{\partial^2 u}{\partial x^2} + \dfrac{\partial^2 u}{\partial y^2} = 0,
\end{equation}

we impose Dirichlet boundary conditions $u(0,y)=\sin(\pi y)$, $u(x,0)=0$, $u(1,y)=0$, $u(x,1)=0$ for $x,y \in [0,1]$. The exact solution to this equation under these conditions is $u(x,y)=e^{-\pi x}\sin(\pi y)$.\newline 

To assess model performance, we evaluate each trained model on a $250$ by $250$ grid, comparing predictions against the analytic solution at each grid point. During training, we resample boundary and collocation points from a uniform distribution at each epoch, using $250$ points for the boundaries and interior points. Each model was trained for $2000$ epochs using the Adam optimizer with a learning rate of $0.02$. We chose this number of epochs as we observed loss convergenced across all models within this time frame.\newline

For this experiment, we adapted the reuploader model described earlier (Figure \ref{circ:student_teacher_circuit}) by enhancing the ansatz layers, using pairs of $R_y$ and $R_x$ gates to increase expressiveness, while feature map layers remained as $R_x$ gates. We compared both CDL and growing circuit methods, training models with reuploader depths ranging from $5$ to $13$ layers.\newline 

The results, displayed in Figure \ref{fig:boxplot_2d_laplace} show the $L_2$ relative error of each model’s predictions on the test grid. Across all layer counts, the growing models consistently outperformed the CDL models, with particularly noticeable degradation in performance for the CDL model at $13$ layers. This performance drop suggests the onset of a barren plateau, characterized by vanishing gradients that hinder effective training in deep quantum circuits.\newline 

Notably, each of the growing models exhibit nearly identical performance and low variance in final error values, indicating that gradual circuit growth leads to more accurate and stable solutions, even in complex, real-world scenarios. This experiment highlights that the growing circuit approach not only mitigates overfitting in artificial tasks but also enhances solution accuracy and robustness in physically significant problems, making it a viable approach for quantum machine learning in PDE contexts.

\section{Conclusion}

This study presents an analysis of training methods for PQCs with a focus on dynamically growing circuits during training. By exploring student-teacher setups where PQCs of matching structure learn the outputs of a randomly initialized teacher circuit. We demonstrated the effectiveness of growth-based training methods across various tasks, including both synthetic and physical systems. Our results reveal that the growing circuit approach outperforms traditional, static methods in several key areas, including accuracy, robustness, and resistance to overfitting, even as circuit depth and model complexity increase.\newline 

For single and two-qubit circuits, the growing methods consistently achieved lower MSE compared to the CDL baseline, with reduced variance in outcomes across random seeds. This stability underscores the potential of circuit growth to provide more reliable convergence, particularly in scenarios where high accuracy is critical. Furthermore, these methods demonstrated an inherent regularizing effect in tasks where noise was introduced, effectively mitigating overfitting by controlling the accessible frequency spectrum during training. This controlled complexity allowed the model to focus on capturing the true underlying patterns in the data, even in the presence of noise, and suggests a promising direction for PQCs in more challenging, noisy environments.\newline 

In a more complex application, solving the $2$D Laplace equation, we observed that the growing PQCs not only achieved lower error than the CDL models but also showed resilience against trainability issues at larger depths. This indicates that dynamically increasing circuit depth offers a scalable solution for quantum machine learning applications, particularly in solving PDEs where model stability and accuracy are essential. Such adaptability may be crucial for extending PQCs to more complex quantum simulations and real-world physical systems, where maintaining expressive yet trainable models is vital.\newline 

Future work could investigate the scalability of these growing methods on even larger PQCs and explore their effectiveness across a wider range of physical and data-driven tasks. Additionally, further studies into adaptive growth schedules—potentially triggered by plateauing validation metrics or gradient signals could yield even more refined control over model complexity. Another promising direction would involve combining circuit growth with advanced regularization techniques to manage both model expressiveness and overfitting, particularly in high-noise quantum machine learning environments.\newline 

In conclusion, our findings establish dynamically growing PQCs as a promising approach for enhancing both the robustness and accuracy of quantum machine learning models, with potential implications for scalable, high-accuracy applications in quantum computing.

\clearpage

\printbibliography

@article{Kyriienko_2021,  
	year = 2021,
	%month = {may},
  
	publisher = {American Physical Society ({APS})},
  
	volume = {103},
  
	number = {5},
  
	author = {Oleksandr Kyriienko and Annie E. Paine and Vincent E. Elfving},
  
%	title = {Solving nonlinear differential equations with differentiable quantum circuits},
  
	journal = {Physical Review A}
}

@article{Grant_2019,
  
	year = 2019,
	%month = {dec},
  
	publisher = {Verein zur Forderung des Open Access Publizierens in den Quantenwissenschaften},
  
	volume = {3},
  
	pages = {214},
  
	author = {Edward Grant and Leonard Wossnig and Mateusz Ostaszewski and Marcello Benedetti},
  
	%title = {An initialization strategy for addressing barren plateaus in parametrized quantum circuits},
  
	journal = {Quantum}
}

@misc{duong2022quantum,
      title={Quantum Neural Architecture Search with Quantum Circuits Metric and Bayesian Optimization}, 
      author={Trong Duong and Sang T. Truong and Minh Tam and Bao Bach and Ju-Young Ryu and June-Koo Kevin Rhee},
      year={2022},
      eprint={2206.14115},
      archivePrefix={arXiv},
      primaryClass={quant-ph}
}

@article{lourensHierarchicalQuantumCircuit2023,
  title = {Hierarchical Quantum Circuit Representations for Neural Architecture Search},
  author = {Lourens, Matt and Sinayskiy, Ilya and Park, Daniel K. and Blank, Carsten and Petruccione, Francesco},
  date = {2023-08-05},
  journaltitle = {npj Quantum Information},
  shortjournal = {npj Quantum Inf},
  volume = {9},
  number = {1},
  pages = {1--15},
  publisher = {{Nature Publishing Group}},
  issn = {2056-6387},
  doi = {10.1038/s41534-023-00747-z},
  issue = {1},
  langid = {english},
}

@article{Schuld_2021,
   title={Effect of data encoding on the expressive power of variational quantum-machine-learning models},
   volume={103},
   ISSN={2469-9934},
   DOI={10.1103/physreva.103.032430},
   number={3},
   journal={Physical Review A},
   publisher={American Physical Society (APS)},
   author={Schuld, Maria and Sweke, Ryan and Meyer, Johannes Jakob},
   year={2021},
   month=mar }

@article{Evci2022GradMaxGN,
  title={GradMax: Growing Neural Networks using Gradient Information},
  author={Utku Evci and Max Vladymyrov and Thomas Unterthiner and Bart van Merrienboer and Fabian Pedregosa},
  journal={ArXiv},
  year={2022},
  volume={abs/2201.05125},
}

@article{Grant2019initialization,
  doi = {10.22331/q-2019-12-09-214},
  url = {https://doi.org/10.22331/q-2019-12-09-214},
  title = {An initialization strategy for addressing barren plateaus in parametrized quantum circuits},
  author = {Grant, Edward and Wossnig, Leonard and Ostaszewski, Mateusz and Benedetti, Marcello},
  journal = {{Quantum}},
  issn = {2521-327X},
  publisher = {{Verein zur F{\"{o}}rderung des Open Access Publizierens in den Quantenwissenschaften}},
  volume = {3},
  pages = {214},
  year = {2019}
}

@misc{fridovichkeil2022spectralbiaspracticerole,
      title={Spectral Bias in Practice: The Role of Function Frequency in Generalization}, 
      author={Sara Fridovich-Keil and Raphael Gontijo-Lopes and Rebecca Roelofs},
      year={2022},
      eprint={2110.02424},
      archivePrefix={arXiv},
      primaryClass={cs.LG},
      url={https://arxiv.org/abs/2110.02424}, 
}

@misc{george2024incrementalspatialspectrallearning,
      title={Incremental Spatial and Spectral Learning of Neural Operators for Solving Large-Scale PDEs}, 
      author={Robert Joseph George and Jiawei Zhao and Jean Kossaifi and Zongyi Li and Anima Anandkumar},
      year={2024},
      eprint={2211.15188},
      archivePrefix={arXiv},
      primaryClass={cs.LG},
      url={https://arxiv.org/abs/2211.15188}, 
}

@article{Skolik_2021,
   title={Layerwise learning for quantum neural networks},
   volume={3},
   ISSN={2524-4914},
   url={http://dx.doi.org/10.1007/s42484-020-00036-4},
   DOI={10.1007/s42484-020-00036-4},
   number={1},
   journal={Quantum Machine Intelligence},
   publisher={Springer Science and Business Media LLC},
   author={Skolik, Andrea and McClean, Jarrod R. and Mohseni, Masoud and van der Smagt, Patrick and Leib, Martin},
   year={2021},
   month=jan }

@misc{zhang2022escapingbarrenplateaugaussian,
      title={Escaping from the Barren Plateau via Gaussian Initializations in Deep Variational Quantum Circuits}, 
      author={Kaining Zhang and Liu Liu and Min-Hsiu Hsieh and Dacheng Tao},
      year={2022},
      eprint={2203.09376},
      archivePrefix={arXiv},
      primaryClass={quant-ph},
      url={https://arxiv.org/abs/2203.09376}, 
}

@article{Holmes_2022,
   title={Connecting Ansatz Expressibility to Gradient Magnitudes and Barren Plateaus},
   volume={3},
   ISSN={2691-3399},
   url={http://dx.doi.org/10.1103/PRXQuantum.3.010313},
   DOI={10.1103/prxquantum.3.010313},
   number={1},
   journal={PRX Quantum},
   publisher={American Physical Society (APS)},
   author={Holmes, Zoë and Sharma, Kunal and Cerezo, M. and Coles, Patrick J.},
   year={2022},
   month=jan }

@article{PRXQuantum.2.020310,
  title = {Qubit-ADAPT-VQE: An Adaptive Algorithm for Constructing Hardware-Efficient Ans\"atze on a Quantum Processor},
  author = {Tang, Ho Lun and Shkolnikov, V.O. and Barron, George S. and Grimsley, Harper R. and Mayhall, Nicholas J. and Barnes, Edwin and Economou, Sophia E.},
  journal = {PRX Quantum},
  volume = {2},
  issue = {2},
  pages = {020310},
  numpages = {16},
  year = {2021},
  publisher = {American Physical Society},
  doi = {10.1103/PRXQuantum.2.020310},
  url = {https://link.aps.org/doi/10.1103/PRXQuantum.2.020310}
}

@article{Du_2022,
   title={Quantum circuit architecture search for variational quantum algorithms},
   volume={8},
   ISSN={2056-6387},
   url={http://dx.doi.org/10.1038/s41534-022-00570-y},
   DOI={10.1038/s41534-022-00570-y},
   number={1},
   journal={npj Quantum Information},
   publisher={Springer Science and Business Media LLC},
   author={Du, Yuxuan and Huang, Tao and You, Shan and Hsieh, Min-Hsiu and Tao, Dacheng},
   year={2022},
   month=may }

@misc{farhi2014quantumapproximateoptimizationalgorithm,
      title={A Quantum Approximate Optimization Algorithm}, 
      author={Edward Farhi and Jeffrey Goldstone and Sam Gutmann},
      year={2014},
      eprint={1411.4028},
      archivePrefix={arXiv},
      primaryClass={quant-ph},
      url={https://arxiv.org/abs/1411.4028}, 
}

@misc{verdon2019learninglearnquantumneural,
      title={Learning to learn with quantum neural networks via classical neural networks}, 
      author={Guillaume Verdon and Michael Broughton and Jarrod R. McClean and Kevin J. Sung and Ryan Babbush and Zhang Jiang and Hartmut Neven and Masoud Mohseni},
      year={2019},
      eprint={1907.05415},
      archivePrefix={arXiv},
      primaryClass={quant-ph},
      url={https://arxiv.org/abs/1907.05415}, 
}

@article{Hadfield_2019,
   title={From the Quantum Approximate Optimization Algorithm to a Quantum Alternating Operator Ansatz},
   volume={12},
   ISSN={1999-4893},
   url={http://dx.doi.org/10.3390/a12020034},
   DOI={10.3390/a12020034},
   number={2},
   journal={Algorithms},
   publisher={MDPI AG},
   author={Hadfield, Stuart and Wang, Zhihui and O’Gorman, Bryan and Rieffel, Eleanor G. and Venturelli, Davide and Biswas, Rupak},
   year={2019},
   month=feb, pages={34} }

@article{PhysRevA.97.022304,
  title = {Quantum approximate optimization algorithm for MaxCut: A fermionic view},
  author = {Wang, Zhihui and Hadfield, Stuart and Jiang, Zhang and Rieffel, Eleanor G.},
  journal = {Phys. Rev. A},
  volume = {97},
  issue = {2},
  pages = {022304},
  numpages = {11},
  year = {2018},
  publisher = {American Physical Society},
  doi = {10.1103/PhysRevA.97.022304},
  url = {https://link.aps.org/doi/10.1103/PhysRevA.97.022304}
}

@article{Streif_2020,
doi = {10.1088/2058-9565/ab8c2b},
url = {https://dx.doi.org/10.1088/2058-9565/ab8c2b},
year = {2020},
publisher = {IOP Publishing},
volume = {5},
number = {3},
pages = {034008},
author = {Michael Streif and Martin Leib},
title = {Training the quantum approximate optimization algorithm without access to a quantum processing unit},
journal = {Quantum Science and Technology},
}

@article{Preskill2018quantumcomputingin,
  doi = {10.22331/q-2018-08-06-79},
  url = {https://doi.org/10.22331/q-2018-08-06-79},
  title = {Quantum {C}omputing in the {NISQ} era and beyond},
  author = {Preskill, John},
  journal = {{Quantum}},
  issn = {2521-327X},
  publisher = {{Verein zur F{\"{o}}rderung des Open Access Publizierens in den Quantenwissenschaften}},
  volume = {2},
  pages = {79},
  month = aug,
  year = {2018}
}

@article{McClean2018BarrenPI,
  title={Barren plateaus in quantum neural network training landscapes},
  author={Jarrod R. McClean and Sergio Boixo and Vadim N. Smelyanskiy and Ryan Babbush and Hartmut Neven},
  journal={Nature Communications},
  year={2018},
  volume={9},
  url={https://api.semanticscholar.org/CorpusID:4465524}
}

@article{PhysRevA.102.032420,
  title = {Robust data encodings for quantum classifiers},
  author = {LaRose, Ryan and Coyle, Brian},
  journal = {Phys. Rev. A},
  volume = {102},
  issue = {3},
  pages = {032420},
  numpages = {24},
  year = {2020},
  publisher = {American Physical Society},
  doi = {10.1103/PhysRevA.102.032420},
  url = {https://link.aps.org/doi/10.1103/PhysRevA.102.032420}
}

@article{PhysRevA.109.042421,
  title = {Let quantum neural networks choose their own frequencies},
  author = {Jaderberg, Ben and Gentile, Antonio A. and Berrada, Youssef Achari and Shishenina, Elvira and Elfving, Vincent E.},
  journal = {Phys. Rev. A},
  volume = {109},
  issue = {4},
  pages = {042421},
  numpages = {10},
  year = {2024},
  publisher = {American Physical Society},
  doi = {10.1103/PhysRevA.109.042421},
  url = {https://link.aps.org/doi/10.1103/PhysRevA.109.042421}
}

\appendix
\onecolumn
\section{\centering Student Teacher}\label{appx:1}

The tables display the performances of models on the student-teacher datasets from section \ref{sec:3.1}, a regression problem on truncated Fourier series. The tables report the mean and standard error of the best losses across the $50$ runs, along with the best and worst of the runs.

\subsection{\centering 1-qubit}\label{appx:11}

The numerical results for the $1$D student-teacher dataset accompanying Figure \ref{fig:boxplot_1d_student_teacher}.

\begin{table*}[h]
    \centering 
    \begin{tabular}{| l | l | l | l |}
    \hline
    \multicolumn{1}{|c|}{}& \multicolumn{3}{|c|}{Mean Squared Error} \\
    \hline
    Model & Mean & Best & Worst\\
    \hline
    5-layer CDL (RAND) & $(1.0\pm4.9)\times 10^{-2}$ & $1.76\times 10^{-6}$ & $2.65\times 10^{-1}$ \\
    
    20-layer CDL (RAND) & $(8.13\pm1.42)\times 10^{-5}$ & $7.63\times 10^{-6}$ & $1.0\times 10^{-2}$ \\ 

    5-layers CDL ($\mathbb{I}$) & $(4.8\pm1.3)\times 10^{-2}$& $1.86\times 10^{-9}$ & $4.35\times 10^{-1}$\\

    20-layer CDL ($\mathbb{I}$) & $(5.5\pm1.1)\times 10^{-5}$& $1.084\times 10^{-6}$ & $3.37\times 10^{-4}$\\

    Block growth & $(1.03\pm3.52)\times 10^{-5}$& $1.46\times 10^{-7}$ & $2.16\times 10^{-4}$\\

    Seq FM Growth & $(1.57\pm5.29)\times 10^{-5}$& $1.060\times 10^{-9}$ & $2.87\times 10^{-4}$\\

    Int FM Growth & $\boldsymbol{(1.029\pm3.52)\times 10^{-6}}$& $\boldsymbol{8.889\times 10^{-10}}$ & $\boldsymbol{1.259\times 10^{-5}}$\\
    \hline
    \end{tabular}
    \caption{Mean squared error (MSE) performance of various models in the $1$D student-teacher task. The table compares the mean MSE (with standard error), as well as the best and worst MSE values achieved over $50$ training runs for each model. Models include complete-depth learning (CDL) approaches with random (RAND) and identity ($\mathbb{I}$) initializations, alongside growth-based methods (Block Growth, Sequential Feature Map Growth, and Interleave Feature Map Growth).}
    \label{table:student_teacher_1d}
\end{table*}

\subsection{\centering 2-qubit}\label{appx:12}

The numerical results for the $2$D student-teacher dataset accompanying Figure \ref{fig:boxplot_2d_student_teacher}.

\begin{table}[h]
    \centering 
    \begin{tabular}{| l | l | l | l |}
    \hline
    \multicolumn{1}{|c|}{}& \multicolumn{3}{|c|}{Mean Squared Error} \\
    \hline
    Model & Mean & Best & Worst\\
    \hline
    5-layer CDL (RAND) & $(5.68\pm5.63)\times10^{-3}$ & $4.47\times 10^{-4}$ & $2.05\times 10^{-2}$ \\
    9-layer CDL (RAND) & $(2.04\pm1.40)\times 10^{-3}$ & $1.75\times 10^{-4}$ & $7.30\times 10^{-3}$ \\ 
    5-layers CDL ($\mathbb{I}$) & $(9.17\pm7.08)\times 10^{-4}$ & $2.83\times 10^{-4}$ & $3.22\times 10^{-3}$\\
    9-layer CDL ($\mathbb{I}$) & $(3.97\pm2.16)\times 10^{-4}$ & $1.44\times 10^{-4}$ & $1.38\times 10^{-3}$\\
    Block Growth & $\boldsymbol{(3.19\pm1.58)\times 10^{-4}}$ & $9.35\times 10^{-5}$ & $1.03\times 10^{-3}$\\
    Seq FM Growth & $(3.21\pm1.39)\times 10^{-4}$ & $\boldsymbol{8.39\times 10^{-5}}$ & $\boldsymbol{7.44\times 10^{-4}}$\\
    Int FM Growth & $(3.75\pm2.98)\times 10^{-4}$ & $1.27\times 10^{-4}$ & $1.97\times 10^{-3}$\\
    \hline 
    \end{tabular}
    \caption{Mean squared error (MSE) performance of various models in the $2$D student-teacher task. The table compares the mean MSE (with standard error), as well as the best and worst MSE values achieved over $50$ training runs for each model. Models include complete-depth learning (CDL) approaches with random (RAND) and identity ($\mathbb{I}$) initializations, alongside growth-based methods (Block Growth, Sequential Feature Map Growth, and Interleave Feature Map Growth).}
    \label{table:student_teacher_2d}
\end{table}

\section{\centering Noisy Student-Teacher}

The numerical results for the $1$D noisy student-teacher dataset accompanying Figure \ref{fig:noisy_student_teacher_losses.pdf}.

\begin{table}[h]
    \centering 
    \begin{tabular}{| l | l | l | l |}
    \hline
    \multicolumn{1}{|c|}{}& \multicolumn{3}{|c|}{Mean Squared Error} \\
    \hline
    Model & Mean & Best & Worst\\
    \hline
    $20$-layer CDL ($\mathbb{I}$)& $(1.58\pm 0.74)\times 10^{-1}$ & $5.95\times 10^{-2}$ & $4.55\times 10^{-1}$ \\ 
    Block Growth & $\boldsymbol{(6.82\pm1.88)\times 10^{-2}}$ & $\boldsymbol{2.34\times 10^{-2}}$ & $\boldsymbol{1.21\times 10^{-1}}$\\
    \hline 
    \end{tabular}
    \caption{Mean squared error (MSE) performance of various models in the $1$D noisy student-teacher task for the test set. The table compares the mean MSE (with standard error), as well as the best and worst MSE values achieved over $50$ training runs for each model. Models include a $20$-layer CDL identity ($\mathbb{I}$) initialized PQC and a block growth PQC.}
    \label{table:student_teacher_noisy_1d}
\end{table}

\end{document}